\documentclass[conference]{IEEEtran}
\usepackage{cite}

\ifCLASSINFOpdf
\else
  \usepackage[dvips]{graphicx}
\fi
\usepackage{color}

\usepackage[cmex10]{amsmath}
\usepackage{amssymb}
\usepackage{multirow}
\allowdisplaybreaks[4]

\usepackage[tight,footnotesize]{subfigure}
\newcommand{\ignore}[1]{}

\begin{document}
%
\title{Analytical Evaluation of Coverage-Oriented Femtocell Network Deployment}




\author{\IEEEauthorblockN{He~Wang\IEEEauthorrefmark{1}\IEEEauthorrefmark{2},
Xiangyun~Zhou\IEEEauthorrefmark{1} and
Mark~C.~Reed\IEEEauthorrefmark{3}\IEEEauthorrefmark{1}}
\IEEEauthorblockA{\IEEEauthorrefmark{1}Research School of Engineering, the Australian National University, ACT 0200, Australia}
\IEEEauthorblockA{\IEEEauthorrefmark{2}\ignore{Canberra Research Laboratory, }National ICT Australia (NICTA), ACT 2601, Australia}
\IEEEauthorblockA{\IEEEauthorrefmark{3}UNSW Canberra, ACT 2600, Australia}
Email: \{he.wang, xiangyun.zhou\}@anu.edu.au, mark.reed@unsw.edu.au}


\maketitle

\begin{abstract}
  This paper proposes a coverage-oriented femtocell network deployment scheme, in which the femtocell base stations (BSs) can decide whether to be active or inactive depending on their distances from the macrocell BSs. Specifically, as the areas close to the macrocell BSs already have satisfactory cellular coverage, the femtocell BSs located inside such areas are kept to be inactive. Thus, all the active femtocells are located in the poor macrocell coverage areas. Based on a stochastic geometric framework, the coverage probability can be analyzed with tractable results. Surprisingly, the results show that the proposed scheme, although with a lower defacto femtocell density, can achieve better coverage performance than that keeping all femtocells in the entire network to be active. The analytical results further identify the achievable optimal performance of the new scheme, which provides mobile operators a guideline for femtocell deployment and operation.
\end{abstract}


%
\IEEEpeerreviewmaketitle

\section{Introduction}\label{sec:Introduction}

In recent years, the mobile communications industry has experienced an unprecedented growth in the numbers of subscribers and data consumption, and the service providers demand a more flexible and advanced cellular network topology different from the traditional one using a macro-centric planning \cite{Qualcomm11LTE_Whitepaper}. Heterogeneous networks, consisting of a diverse set of wireless technologies, including macrocell base stations (BSs) and low-power access points, can be deployed to most efficiently use the dimensions of space and frequency \cite{SinDhi12arXiv}. Being an important part of heterogeneous networks, femtocell access points (or called femtocell BSs) are user-deployed low-power devices operating in the licensed spectrum \cite{Sau09Book, ChaAnd08MCOM}. \ignore{Backhauled onto the operator's network via the IP-based wired connection\ignore{, such as digital subscriber line (DSL), cable broadband access, or fiber}, femtocells are designed to provide voice and high data-rate sustained services for indoor environments, where a majority of user traffic comes from.}By off-loading indoor traffic from the macrocells to femtocells and decreasing the distance from users to BSs, femtocells bring a multitude of benefits, including improved user experiences and more efficient spatial reuse of spectrum \cite{AndCla12JSAC}.

However, the randomness of femtocell BSs' locations introduced by their user-deployed nature presents us with a design challenge on improving the cellular coverage. With a constant pre-configurable transmit power, which is a mode commonly implemented in current solutions \cite{ChaKou09TWC, WeiGro10MCOM}, the femtocell coverage range is significantly reduced when it is close to a macrocell or picocell BS site \cite{JoSan12TWC}. More interestingly, due to the completely random deployment of femtocell BSs, increasing the density of femtocell BSs does not give any noticeable improvement in the coverage probability \cite{JoSan12TWC, DhiGan12JSAC, WanRee12AusCTW}. The main cause of this fact is the increased network interference from having more femtocell BSs in satisfactory macrocell areas. Hence, one interesting question raised from the above discussion is whether we can improve the cellular coverage by deactivating some of the femtocell BSs at undesirable locations.

\begin{figure}[tb!]
  \centering
  \includegraphics[width=0.35\textwidth, bb=140 262 443 563, clip = true]{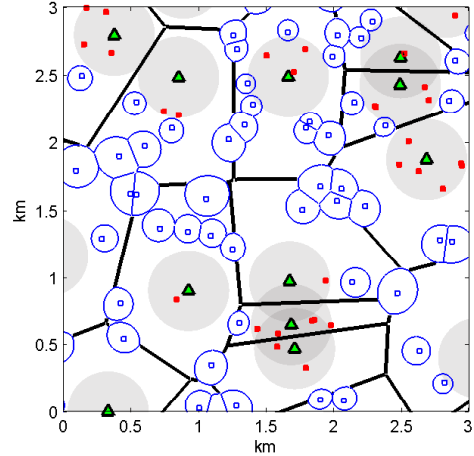}
  \vspace{-0.18in}
  \caption{Coverage regions for a two-tier network with coverage-oriented femtocell deployment. Both macrocell BSs (triangles) and femtocell BSs (squares) are distributed as independent PPPs with $P_1/P_2 = 26$ dB and $\lambda_2/\lambda_1 = 10$. Femtocell BSs are inactive (marked as red filled squares) in the inner regions (shadow areas), while femtocell BSs are active (marked as blue nonfilled squares) in the outer regions.} \label{fig:CoverageOriented_Deployment_Demo}
\end{figure}

Therefore, in this work, we consider a simple and straightforward idea: avoiding using femtocell BSs in the regions where macrocell coverage is already satisfactory, which are usually the areas around all macrocell BSs and defined as the \emph{inner regions}. As illustrated in Fig. \ref{fig:CoverageOriented_Deployment_Demo}, all the femtocell BSs located within a certain distance away from any macrocell sites remain inactive, marked as red filled squares therein. This scheme guarantees that femtocell BSs are deployed in a coverage-oriented fashion and the defacto overall femtocell density is decreased. Furthermore, this coverage-oriented deployment scheme can be achieved easily in practice, by implementing a listening model to the downlink signal level from the strongest BS \cite{YavMes09MCOM} to estimate the distance and then determining the femtocell BS's on/off status. We aim to show the positive impact of employing the proposed scheme on the downlink coverage performance of the heterogeneous cellular networks.

\subsection{Approach and Contributions}
Fortunately, modelling BSs to be randomly placed points in a plane and utilising stochastic geometry \cite{StoKen95Book}\cite{BacBla09Book} to study cellular networks has been used extensively as an analytical tool for improving tractability \cite{Bro00JSAC, YanPet03TSP, Hae08TIT}. Recent works \cite{SinDhi12arXiv, AndBac11JCOM, NovGan11TWC, JoSan12TWC, DhiGan12JSAC} have shown: compared to the practical network deployment, modelling the cellular network with BS locations drawn from a homogeneous Poisson Point Process (PPP) is as accurate as the traditional grid models. More importantly, the stochastic geometry model can provide more tractable analytical results on the coverage and throughput performance. For these reasons we also adopt PPPs to model the locations of BSs of the cellular networks in this paper.

The main contributions of this paper are as follows:
\begin{enumerate}
  \item A simple coverage-oriented femtocell deployment scheme is proposed. In this scheme, femtocell BSs within a certain distance away from any macrocell BSs are automatically turned to be inactive. By employing this, we can guarantee that all the active femtocell BSs are deployed in the areas with unsatisfied macrocell coverage.
  \item We provide the tractable probabilistic characterisation of the coverage probability at a randomly located mobile user in the new scheme.
  \item Our analytical results clearly quantify the impact of inner region size on the cellular coverage probability. For networks adopting the proposed scheme, the service providers are able to use the results in this work to carefully set the condition of femtocell deactivation for optimal coverage performance.
\end{enumerate}

The remainder of the paper is organised as follows. In Section \ref{sec:SysModel} network model and connectivity model used in this work are introduced. Section \ref{sec:CovDeployment} provides the tractable result for the coverage performance of the proposed coverage-oriented femtocell deployment scheme. Section \ref{sec:NumResults} presents numerical results and we conclude the paper in Section \ref{sec:Conclusion}.

\section{System Model}\label{sec:SysModel}
\subsection{Two-Tier Cellular Network Model}
A heterogeneous cellular network consisting of two tiers, i.e., macrocell and femtocell networks, is assumed in this analysis, where spatial deployment densities, transmit powers may be different across tiers. Specifically, BSs in the macrocell and femtocell tiers are spatially distributed as a two-dimensional (2-D) PPP $\Phi_{1}$ and $\Phi_{2}$ with the densities of $\lambda_{1}$ and $\lambda_{2}$ respectively. The macrocell and femtocell BSs have the transmit power $P_{tx,1}$ and $P_{tx,2}$ respectively, and the same transmit power holds across each tier. We consider an independent collection of mobile users in this paper, whose locations are according to some independent stationary point process. Without any loss of generality the mobile user under analysis is assumed located at the origin.

Here we use the standard power loss propagation model with path loss exponent $\alpha > 2$ and path loss constant $L_0$ at the reference distance $r_0 = 1$m. We assume that the typical mobile user experiences Rayleigh fading from the serving and all the interfering BSs. The fading's impact on the signal power is denoted by the random variable $h$, which follows an exponential distribution with mean $1$, i.e. $h \sim \exp(1)$. The noise power is assumed to be additive and constant with value $\sigma^2$.

\subsection{Maximum Long-term Received Power Connectivity Model}
As a widely used special case of the general cell association model in \cite{JoSan12TWC}, the mobile user intuitively connects to the BS providing maximum long-term received power. Specifically, $\omega$ is the selected tier number and can be given as,
  \begin{equation}\label{eqn:omega}
    \omega = \arg \max_{i \in \mathbb{K}} \big[ {P_{tx,i} L_0 (r_{min,i})^{-\alpha}} \big],
  \end{equation}
in which $\mathbb{K} \triangleq \{1,2\}$ and $r_{min,i}$ is the minimum distance from the $i$-th tier BSs to the typical mobile user at the origin, i.e., $r_{min,i} = \min_{x \in \Phi_{i}} \|x\|$.
For simplicity, we use $P_i$ to denote the product of $P_{tx,i}$ and $L_0$, i.e., $P_i \triangleq P_{tx,i} L_0, i \in \mathbb{K}$. The random variable $q$ is defined as the maximum long-term received power from all BSs,
  \begin{equation}\label{eqn:q}
    q  \triangleq  {P_{\omega} (r_{min,\omega})^{-\alpha}}.
  \end{equation}

If the typical user utilises the above connectivity model to select the serving cell, its downlink received SINR can be expressed as
  \begin{equation}\label{eqn:SINR}
    \mathrm{SINR} = \frac{P_\omega h (r_{min, \omega})^{-\alpha}}{I_q + \sigma^2},
  \end{equation}
where $I_q$ is the cumulative interference from all other BSs (except the BS $b_o$ from $\omega$-th tier serving for the mobile user at $o$), i.e.,
  \begin{equation}\label{eqn:Interference}
    I_q = \sum_{i \in \mathbb{K}} \sum_{x \in \Phi_i \setminus \{b_o\}} P_i  h_{x} \|x\|^{-\alpha}.
  \end{equation}
In our study, no intracell interference is incorporated since we suppose the orthogonal multiple access employed among all users served by the same cell.

The coverage probability is defined as $p_{c} (T) \triangleq \mathbb{P}[\mathrm{SINR} > T ]$, i.e., the probability of a target SINR $T$ (or SINR threshold) achievable at the typical user. This probability of coverage is also exactly the complementary cumulative distribution function (CCDF) of SINR over the entire network.

\subsection{Coverage Probability for Uniform Femtocell Deployment}
Following the analysis in \cite{JoSan12TWC}, we firstly provide the result for the traditional uniform femtocell deployment, where all the PPP-distributed femtocell BSs stay active. In this case, the cumulative distribution function (CDF) of the random variable $q$ can be derived as
  \begin{eqnarray}\label{eqn:q_cdf}
    F_q(t) \ignore{&=& \mathbb{P}\big[ {P_{\omega} (r_{min,\omega})^{-\alpha}} \leqslant t \big] \nonumber \\}
    &=& \prod_{i \in \mathbb{K}} \mathbb{P}\big[ {P_{i} (r_{min,i})^{-\alpha}} \leqslant t \big] \nonumber \\
    \ignore{&=& \prod_{i \in \mathbb{K}} \mathbb{P}\big[ r_{min,i} \geqslant \Big(\frac{P_i}{t}\Big)^{1/\alpha} \big] \nonumber \\}
    &\stackrel{(a)}{=}& \prod_{i \in \mathbb{K}} \exp \big[- \pi \lambda_i \Big(\frac{P_i}{t} \Big)^{2/\alpha} \big] \nonumber \\
    &=& \exp \big(- \pi \sum_{i \in \mathbb{K}} {\lambda_i P_i^{2/\alpha}} t^{-2/\alpha}\big),
  \end{eqnarray}
where $(a)$ follows that the null probability of a 2-D homogeneous PPP with density $\lambda$ in an area $A$ is $\exp(-\lambda A)$ \cite{StoKen95Book}. Next, the probability density function (pdf) of the random variable $q$ can be provided as
  \begin{eqnarray}\label{eqn:q_pdf}
    f_q(t) = \frac{\mathrm{d} F_q(t)}{\mathrm{d} t}=   \frac{2 \pi \xi}{\alpha} (t^{-\frac{2}{\alpha}-1}) \exp(- \pi \xi t^{-\frac{2}{\alpha}}),
  \end{eqnarray}
where $\xi$ is a constant value depending on the transmit powers ($P_i$ for the $i$-th tier) and the BS densities ($\lambda_i$ for the $i$-th tier), i.e.,
  \begin{equation}\label{eqn:xi}
    \xi = \sum_{i \in \mathbb{K}} {\lambda_i P_i^{2/\alpha}}.
  \end{equation}
The coverage probability at the typical user is
  \begin{align}\label{eqn:probcov}
    p_{c} (T) &= \mathbb{P}[\mathrm{SINR} > T ] \nonumber \\
    &=  \int_{t>0} \mathbb{P} [\mathrm{SINR} > T \mid q=t] f_q(t) \mathrm{d}t \nonumber \\
    &=  \int_{t>0} \exp (-\frac{T \sigma^2}{t} ) {\mathcal{L}}_{I_{q}}\big(\frac{T}{t}\mid q = t\big) f_q(t) \mathrm{d}t \nonumber \\
    &=  \frac{2 \pi \xi}{\alpha} \int_{t>0} \exp (-\frac{T \sigma^2}{t} ) \nonumber \\
    & \ \ \ \ \ \ \ \  \cdot \exp \big[- \pi \xi \big(1+\rho(T,\alpha)\big) t^{-2/\alpha}\big] t^{-\frac{2}{\alpha}-1} \mathrm{d}t,
  \end{align}
where ${\mathcal{L}}_{I_{q}}(\cdot \mid q = t)$ is the Laplace transform of random variable $I_q$ and the function $\rho(x,\alpha)$ can be expressed as
  \begin{eqnarray}\label{eqn:rho}
    \rho(x,\alpha) = x^{2/\alpha} \int_{x^{-2/\alpha}}^{\infty} \frac{1}{1+u^{\alpha/2}} \mathrm{d}u.
  \end{eqnarray}

\section{Coverage-Oriented Deployment for Femtocell Tier}\label{sec:CovDeployment}

\subsection{Outer and Inner Regions}
As described in Section \ref{sec:Introduction}, we propose the coverage-oriented femtocell deployment scheme, in which all the femtocell BSs in the inner regions stay inactive and other femtocell BSs in the outer regions remain active.

Specifically, the inner regions are defined as the areas where the distance from the nearest macrocell BS site is larger than $D$, and the outer regions are defined as the areas where the distances from any macrocell BSs are no less than $D$, i.e.,
  \begin{equation}\label{eqn:A}
    A_{inner} = \bigcup_{x \in \Phi_1} B(x,D) \text{ and } A_{outer} = \mathbb{R}^2 \setminus A_{inner},
  \end{equation}
where $D$ is called the radius of inner region in this paper.

\subsection{Maximum Long-term Received Power from the Serving BS} \label{subsec:equRxPwr}
Given the condition of the typical user located in the outer regions $A_{outer}$, we approximate the density of femtocell BSs in the vicinity of the typical user as $\lambda_2$. This approximation is accurate as long as the typical user is not located close to the boundary between the inner and outer regions. Hence, the CDF of the random variable $q$ can be approximated similar to (\ref{eqn:q_cdf}), i.e.,
  \begin{align}\label{eqn:q_outerarea_cdf}
    & F_{q \mid o \in A_{outer}}(t)  \approx \prod_{i \in \mathbb{K}} \mathbb{P}\big[ r_{min,i} \geqslant \big({P_i}/{t}\big)^\frac{1}{\alpha} \mid r_{min,1} \geqslant D\big] \nonumber \\
    & = \begin{cases}
      {\exp(- \pi {\lambda_2 P_2^{\frac{2}{\alpha}}} t^{-\frac{2}{\alpha}})} & \text{if $t > \frac{P_1}{D^{\alpha}} $}\\
      {\exp(- \pi \sum_{i \in \mathbb{K}} {\lambda_i P_i^{\frac{2}{\alpha}}} t^{-\frac{2}{\alpha}})}/{\exp(-\pi \lambda_1 D^2)} & \text{if $t \leqslant \frac{P_1}{D^{\alpha}} $}\\
   \end{cases}.
  \end{align}
Then the pdf of the random variable $q$ for the outer region is derived as
  \begin{align}\label{eqn:q_outerarea_pdf}
    & f_{q \mid o \in A_{outer}}(t) = {\mathrm{d} F_{q \mid o \in A_{outer}}(t)}/{\mathrm{d} t} \nonumber \\
    &\approx \begin{cases}
      \frac{2 \pi}{\alpha} \lambda_2 P_2^{\frac{2}{\alpha}} \exp(- \pi \lambda_2 P_2^{\frac{2}{\alpha}} t^{-\frac{2}{\alpha}}) t^{-\frac{2}{\alpha}-1} & \text{if $t > \frac{P_1}{D^{\alpha}} $}\\
      \frac{2 \pi}{\alpha} \xi \exp(- \pi \xi t^{-\frac{2}{\alpha}}) t^{-\frac{2}{\alpha}-1} /{\exp(-\pi \lambda_1 D^2)} & \text{if $t \leqslant \frac{P_1}{D^{\alpha}} $}\\
   \end{cases}. \nonumber \\
  \end{align}

For the case when the typical user lies in the inner regions $A_{inner}$, this user is served by a femtocell BS with a small probability. By assuming that the typical user always gets service from macrocell BSs, the CDF of the random variable $q$ can be approximated by
  \begin{align}\label{eqn:q_innerarea_cdf}
    & F_{q \mid o \in A_{inner}}(t) \approx \mathbb{P}\big[ r_{min,1} \geqslant \big({P_i}/{t}\big)^\frac{1}{\alpha} \mid r_{min,1} < D \big] \nonumber \\
    &=\begin{cases}
      \frac{\exp (\pi {\lambda_1 P_1^\frac{2}{\alpha}} t^{-\frac{2}{\alpha}}) - \exp(- \pi \lambda_1 D^2) }{1- \exp(- \pi \lambda_1 D^2)} & \text{when $t > \frac{P_1}{D^{\alpha}} $}\\
      0 & \text{when $t \leqslant \frac{P_1}{D^{\alpha}} $}\\
   \end{cases}.
  \end{align}
Then the pdf of the random variable $q$ for the inner region is provided as
  \begin{align}\label{eqn:q_innerarea_pdf}
    & f_q(t \mid o \in A_{inner})  = {\mathrm{d} F_{q \mid o \in A_{inner}}(t)}/{\mathrm{d} t} \nonumber \\
    & \approx  \begin{cases}
      \frac{2 \pi \lambda_1 P_1^{\frac{2}{\alpha}} t^{-\frac{2}{\alpha}-1} \exp(- \pi \lambda_1 P_1^{\frac{2}{\alpha}} t^{-\frac{2}{\alpha}})}{\alpha [1 - \exp(-\pi \lambda_1 D^2) ]} & \text{when $t > \frac{P_1}{D^{\alpha}} $}\\
      0 & \text{when $t \leqslant \frac{P_1}{D^{\alpha}} $}\\
   \end{cases}.
  \end{align}

\subsection{The SINR Distribution for Outer Regions} \label{subsec:covprobouter}

For the typical user in the outer regions, its coverage probability can be given as
  \begin{align}\label{eqn:probcov_outer}
    p_{c,o} (T) &= \mathbb{P}[\mathrm{SINR} > T \mid o \in A_{outer}] \nonumber \\
    &=  \mathbb{E}_{q}\big[\mathbb{P}[\mathrm{SINR} > T \mid q, o \in A_{outer}]\big] \nonumber \\
    &=  \int_{t>0} \mathbb{P} [\mathrm{SINR} > T \mid q=t, o \in A_{outer}] \nonumber \\
    &  \ \ \ \ \ \ \ \ \ \ \ \ \ \ \ \ \ \ \ \ \ \ \ \ \ \ \ \ \cdot f_q(t \mid o \in A_{outer}) \mathrm{d}t \nonumber \\
    &\stackrel{(a)}{=}  \int_{t>0} \exp (-\frac{T \sigma^2}{t} ) {\mathcal{L}}_{I_{q}}\big(\frac{T}{t}\mid q = t, o \in A_{outer}\big) \nonumber \\
    &  \ \ \ \ \ \ \ \ \ \ \ \ \ \ \ \ \ \ \ \ \ \ \ \ \ \ \ \ \cdot f_q(t \mid o \in A_{outer}) \mathrm{d}t,
  \end{align}
where $(a)$ follows from the Rayleigh fading assumption (i.e. $h \sim \exp(1)$), and ${\mathcal{L}}_{I_{q}}(\cdot \mid q = t, o \in A_{outer})$ is the Laplace transform of random variable $I_q$ given the condition that the typical user is located in the outer regions. Assuming the interference from femtocell BSs comes from the whole 2-D plane, we can approximate this Laplace transform based on different conditions, $q = t \leqslant \frac{P_1}{D^{\alpha}}$ and $q = t > \frac{P_1}{D^{\alpha}}$ respectively, i.e.,
  \begin{multline}\label{eqn:laplacetransform_outer_1}
  {\mathcal{L}}_{I_{q}}\big(\frac{T}{t}\mid q = t \leqslant \frac{P_1}{D^{\alpha}}, o \in A_{outer}\big) \\
    \approx \exp \big[-{\pi} \big(\sum_{i \in \mathbb{K}} \lambda_i P_i^{2/\alpha}\big) \rho(T,\alpha) {t^{-2/\alpha}} \big],
  \end{multline}
and
  \begin{eqnarray}\label{eqn:laplacetransform_outer_2}
    & & {\mathcal{L}}_{I_{q}}\big(\frac{T}{t}\mid q = t > \frac{P_1}{D^{\alpha}}, o \in A_{outer}\big) \nonumber \\
    & \approx & \exp \big[-2 \pi \lambda_1 \int_{D}^{\infty} \frac{P_1 T}{P_1 T + t v^{\alpha}} v \mathrm{d}v \big] \nonumber \\
    & & \ \ \ \ \ \ \ \ \ \ \ \ \ \ \ \ \cdot \exp \big[-{\pi} \lambda_2 P_2^{2/\alpha} \rho(T,\alpha) {t^{-2/\alpha}} \big] \nonumber \\
    &=& \exp \big[-\pi \lambda_1 D^2 \rho \big(\frac{P_1 T}{D^{\alpha} t},\alpha \big) \big] \nonumber \\
    & & \ \ \ \ \ \ \ \ \ \ \ \ \ \ \ \ \cdot \exp \big[-{\pi} \lambda_2 P_2^{2/\alpha} \rho(T,\alpha) {t^{-2/\alpha}} \big].
  \end{eqnarray}
It is a pessimistic assumption since the original femtocell interference from inner area is eliminated due to this coverage-oriented deployment scheme, and the numerical results in the next section show that it is still a reasonably accurate approximation. By substituting (\ref{eqn:q_outerarea_pdf}), (\ref{eqn:laplacetransform_outer_1}) and (\ref{eqn:laplacetransform_outer_2}) into (\ref{eqn:probcov_outer}), we can have
  \begin{multline}\label{eqn:probcov_outer_proof}
    p_{c,o} (T) \approx  \frac{2 \pi \xi}{\alpha \exp(- \pi \lambda_1 D^2)} \int_{0}^{\frac{P_1}{D^{\alpha}}} \exp (-\frac{T \sigma^2}{t} ) \\
    \exp \big[-\frac{\pi \xi}{t^{2/\alpha}}\big(1+\rho(T,\alpha)\big) \big] t ^{-\frac{2}{\alpha}-1}\mathrm{d}t \\
    + \frac{2 \pi \lambda_2 P_2^{{2}/{\alpha}}}{\alpha} \int_{\frac{P_1}{D^{\alpha}}}^{\infty} \exp (-\frac{T \sigma^2}{t} )
    \exp \big[-\pi \lambda_1 D^2 \rho(\frac{P_1 T}{D^{\alpha} t},\alpha) \big]\\
    \cdot \exp \big[-{\pi} \lambda_2 P_2^{2/\alpha} \big(1+\rho(T,\alpha)\big) {t^{-2/\alpha}} \big]
    t^{-\frac{2}{\alpha}-1}\mathrm{d}t.
  \end{multline}

\subsection{The SINR Distribution for Inner Regions} \label{subsec:covprobinner}

If the typical user is located in the inner regions, its coverage probability can be given as
  \begin{align}\label{eqn:probcov_inner}
    p_{c,i}(T) &= \mathbb{P}[\mathrm{SINR} > T \mid o \in A_{inner}] \nonumber \\
    &=  \mathbb{E}_{q}\big[\mathbb{P}[\mathrm{SINR} > T \mid q, o \in A_{inner}]\big] \nonumber \\
    &\stackrel{(a)}{=}  \int_{t>0} \exp (-\frac{T \sigma^2}{t} ) {\mathcal{L}}_{I_{q}}\big(\frac{T}{t}\mid q = t, o \in A_{inner}\big) \nonumber \\
    & \ \ \ \ \ \ \ \ \ \ \ \ \ \ \ \ \ \ \ \ \ \ \ \ \ \ \ \ \cdot f_q(t \mid o \in A_{inner}) \mathrm{d}t,
  \end{align}
where $(a)$ still comes from the Rayleigh fading assumption, and ${\mathcal{L}}_{I_{q}}(\cdot \mid q = t, o \in A_{inner})$ is the Laplace transform of random variable $I_q$ given the condition that the typical user is located in the inner region. Here we assume the femtocell interference comes from the whole region out of the area $B(o,D)$, which is an optimistic estimation (proved to be accurate by the numerical results). Then, we can have
  \begin{eqnarray}\label{eqn:laplacetransform_inner}
  & & {\mathcal{L}}_{I_{q}}\big(\frac{T}{t}\mid q = t, o \in A_{inner}\big) \nonumber \\
  & \approx & \exp \big[-{\pi} \lambda_1 P_1^{2/\alpha} \rho(T,\alpha) {t^{-2/\alpha}} \big] \nonumber \\
  & & \ \ \ \ \ \ \ \ \ \ \ \cdot \exp \big[-2 \pi \lambda_2 \int_{D}^{\infty} \frac{P_2 T}{P_2 T + t v^{\alpha}} v \mathrm{d}v \big], \nonumber \\
  & = &  \exp \big[-{\pi} \lambda_1 P_1^{2/\alpha} \rho(T,\alpha) {t^{-2/\alpha}} \big] \nonumber \\
  & & \ \ \ \ \ \ \ \ \ \ \ \cdot \exp \big[-\pi \lambda_2 D^2 \rho(\frac{P_2 T}{D^{\alpha} t},\alpha) \big].
  \end{eqnarray}

By substituting (\ref{eqn:q_innerarea_pdf}) and (\ref{eqn:laplacetransform_inner}) into (\ref{eqn:probcov_inner}), we can obtain
  \begin{multline}\label{eqn:probcov_inner_proof}
    p_{c,i}(T) \approx \frac{2 \pi \lambda_1 P_1^{2/\alpha}}{\alpha [1 - \exp(-\pi \lambda_1 D^2)]} \cdot \\
    \int_{\frac{P_1}{D^{\alpha}}}^{\infty} \exp (-\frac{T \sigma^2}{t}) \exp \big[-{\pi} \lambda_1 P_1^{\frac{2}{\alpha}} \big(1+\rho(T,\alpha)\big) {t^{-\frac{2}{\alpha}}} \big] \\
    \cdot \exp \big[-\pi \lambda_2 D^2 \rho(\frac{P_2 T}{D^{\alpha} t},\alpha) \big] t^{-\frac{2}{\alpha}-1} \mathrm{d}t.
  \end{multline}

\subsection{The Overall Coverage Probability} \label{subsec:covproboverall}

For the typical user, the coverage probability (or equivalently SINR's CCDF) is
  \begin{align}\label{eqn:probcov_overall}
    p_{c}(T) &= p_{c,i}(T) \cdot \mathbb{P}[o \in A_{inner}] + p_{c,o}(T) \cdot \mathbb{P}[o \in A_{outer}] \nonumber \\
    &= p_{c,i}(T) \cdot \big[1-\exp(-\pi \lambda_1 D^2)\big] \nonumber \\
    &   \ \ \ \ \ \ \ \ \ \ \ \ \ \ \ \ \ \ \ \  + p_{c,o}(T) \cdot \big[\exp(-\pi \lambda_1 D^2)\big],
  \end{align}
where $\mathbb{P}[o \in A_{inner}]$ and $\mathbb{P}[o \in A_{outer}]$ are the probabilities of the typical user located in inner and outer regions respectively. An approximation of the coverage probability can hence be obtained by substituting (\ref{eqn:probcov_outer_proof}) for outer regions and (\ref{eqn:probcov_inner_proof}) for inner regions into the above expression.

\section{Numerical Results}\label{sec:NumResults}

\begin{figure}[b!]
  \centering
  \includegraphics[width=0.455\textwidth, bb=108 267 480 575]{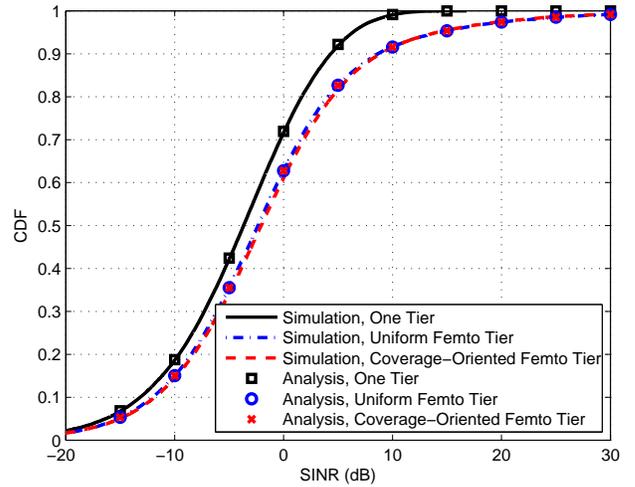}
  \caption{SINR distributions (CDF curves) for outer regions, $D=400$m, and $\lambda_2/\lambda_1 = 10$.} \label{fig:Pc_OuterArea_radius_p2}
\end{figure}

In this section, we present numerical results on the SINR distributions for the proposed coverage-oriented femtocell deployment scheme. Here we assume the transmit powers of macrocell and femtocell BSs as $P_{tx,1} = 46$ dBm and $P_{tx,2} = 20$ dBm respectively. The macrocell tier density is $\lambda_1 = 1$ per square km in all numerical results. The path loss constant and exponent are assumed to be $L_0 = - 34$ dB and $\alpha = 4$. The thermal noise power is $\sigma^2 = -104$ dBm (i.e., $10$ MHz bandwidth). Monte Carlo simulations are also conducted to compare with our analysis for the purpose of model validation.

\begin{figure}[thbp!]
  \centering
  \includegraphics[width=0.455\textwidth, bb=108 267 480 575]{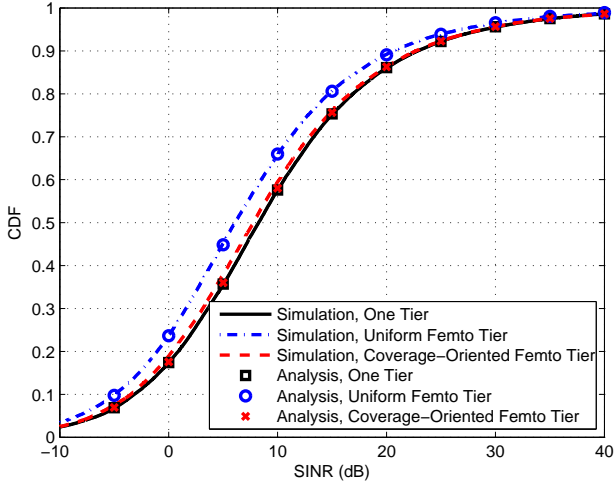}
  \caption{SINR distributions (CDF curves) for inner regions, $D=400$m, and $\lambda_2/\lambda_1 = 10$.} \label{fig:Pc_InnerArea_radius_p2}
\end{figure}

\begin{figure}[thbp!]
  \centering
  \includegraphics[width=0.455\textwidth, bb=108 267 480 575]{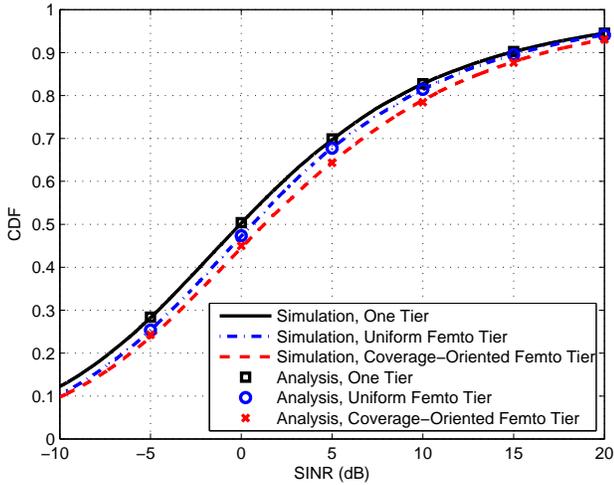}
  \caption{SINR distributions (CDF curves) for the whole area, $D=400$m, and $\lambda_2/\lambda_1 = 10$.} \label{fig:Pc_WholeArea_radius_p2}
\end{figure}

Fig. \ref{fig:Pc_OuterArea_radius_p2} and Fig. \ref{fig:Pc_InnerArea_radius_p2} demonstrate the results of SINR distributions (CDF curves) for outer and inner regions respectively, given the condition of $D=400$m, $\lambda_2/\lambda_1 = 10$. Firstly, the tractable analytical results, i.e., the approximations derived in the previous section, for both outer and inner regions are reasonably accurate. By deploying femtocell in the outer regions, both uniform and coverage-oriented methods can provide significant and nearly the same improvement over the single macrocell tier deployment. This clearly shows the necessity to deploy femtocell BSs in the bad macrocell coverage areas. On the other hand, the inner regions are more likely to have a better macrocell coverage, in which deploying femtocell BSs can only degrade the coverage probability (as shown in the results of the uniform femtocell tier). The new scheme avoids femtocell deployment in the inner regions, which is able to achieve the performance nearly as good as only one macrocell tier deployed. Through combining the results of outer and inner regions by using (\ref{eqn:probcov_overall}), the CDF curves for any randomly chosen users are illustrated in Fig. \ref{fig:Pc_WholeArea_radius_p2}. We can observe from the accurate tractable analysis that the coverage-oriented femtocell deployment scheme outperforms both single and two-tier uniform deployment in terms of coverage performance.

\begin{figure}[t!]
  \centering
  \includegraphics[width=0.455\textwidth, bb=108 267 480 575]{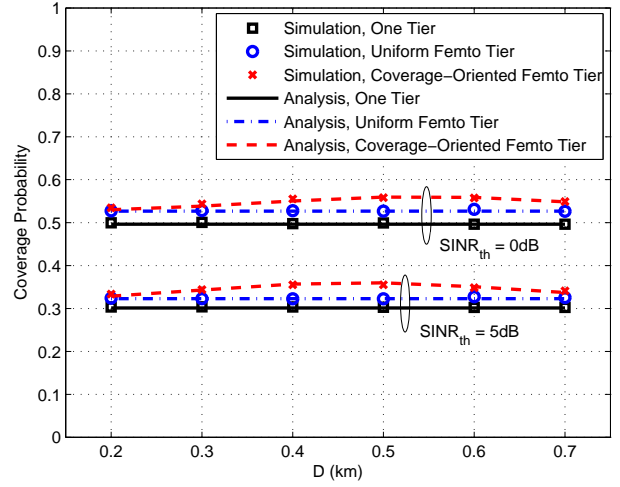}
  \caption{Coverage probability over different $D$, $\lambda_2/\lambda_1 = 10$, SINR thresholds are $0$ dB and $5$ dB respectively.}  \label{fig:CovProb_overradius_lambda1to10}
\end{figure}

\begin{figure}[t!]
  \centering
  \includegraphics[width=0.455\textwidth, bb=108 267 480 575]{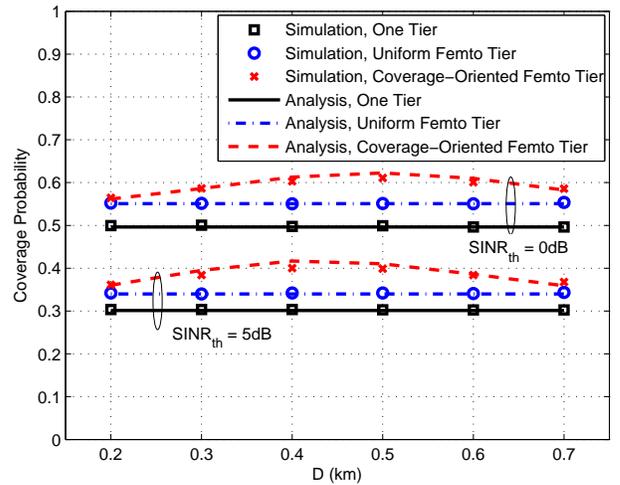}
  \caption{Coverage probability over different $D$, $\lambda_2/\lambda_1 = 40$, SINR thresholds are $0$ dB and $5$ dB respectively.}  \label{fig:CovProb_overradius_lambda1to40}
\end{figure}

Furthermore, by presenting the achievable coverage probability versus the inner region radius $D$ in Fig. \ref{fig:CovProb_overradius_lambda1to10} and Fig. \ref{fig:CovProb_overradius_lambda1to40}, we can see the importance of properly dividing inner and outer regions on the coverage performance. Take the case of $\lambda_2/\lambda_1 = 10$ and SINR threshold to claim outage is $0$ dB shown in Fig. \ref{fig:CovProb_overradius_lambda1to10} for example, the achievable coverage probability is around $56\%$ at $D = 500$ m for the proposed scheme, compared with $53\%$ for uniform two-tier deployment and $50\%$ for single macrocell tier. By increasing the femtocell tier density to $\lambda_2/\lambda_1 = 40$ as illustrated in Fig. \ref{fig:CovProb_overradius_lambda1to40}, the coverage performance for coverage-oriented scheme is enhanced more obviously than traditional uniform deployment, which demonstrates the benefits provided by the proposed new scheme, i.e., $61\%$ for coverage-oriented deployment and $55\%$ for uniform two-tier deployment. From both figures, similar enhancement can be observed if we set the SINR threshold to be $5$ dB. It should be noticed that the slight mismatches between simulation and tractable results in Fig. \ref{fig:CovProb_overradius_lambda1to40} come from the approximation for the inner region analysis, but the performance trend can be well captured by the tractable results. More importantly, the new scheme does not cost any further network resources; on the contrary, we only deactivate some of the femtocell BSs located in the positions not suitable for femtocell deployment.

\begin{figure}[t!]
  \centering
  \includegraphics[width=0.455\textwidth, bb=108 267 480 575]{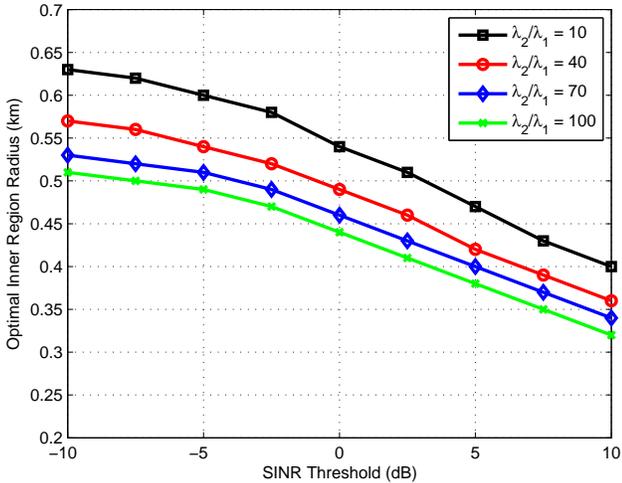}
  \caption{Optimal inner region radius over different SINR thresholds.}  \label{fig:OptimalDoverSINRth}
\end{figure}

The tractable results provide us a way to find the best inner regions radius value to achieve the maximum coverage probability for each set of parameters. As illustrated in Fig. \ref{fig:OptimalDoverSINRth}, we can observe that the optimal $D$ is decreased across both the SINR thresholds and BS density ratios $\lambda_2/\lambda_1$.


\section{Conclusion}\label{sec:Conclusion}
In this work, we studied the coverage performance of the cellular networks with a newly proposed coverage-oriented femtocell deployment scheme. Using the tools from stochastic geometry, tractable results to characterise the coverage probability were obtained. The numerical results validated the tractable expressions and approximations, which provided the following important message: by carefully choosing the parameters for the proposed coverage-oriented femtocell deployment scheme, the coverage performance can be improved at nearly no cost of network resource. Our results can be utilised as guidelines for femtocell deployment to enhance coverage performance.

Future extensions to this analysis could focus on applying more general connectivity model with cell association biasing for different tiers and femtocell deployment over a more complex network topology, such as the one with macrocell and picocell coexisting. The downlink throughput performance with the proposed scheme is also currently under our investigation.


\section*{Acknowledgment}
H. Wang is with the Australian National University and NICTA. NICTA is funded by the Australian Government as represented by the Department of Broadband, Communications and the Digital Economy and the Australian Research Council through the ICT Centre of Excellence program. This work was supported by the Australian Research Council¡¯s Discovery Projects funding scheme (Project No. DP110102548 and Project No. DP130101760).



\vspace{0.55cm}
\bibliographystyle{IEEEtran}
\bibliography{IEEEabrv,Bib_Database}
%
%
%

\end{document}